\documentclass[12pt]{article}
\usepackage{amssymb,amsmath,slashed,latexsym,float}
\usepackage{float}
\allowdisplaybreaks
\newcommand{\ft}[2]{{\textstyle\frac{#1}{#2}}}
\def\Re{\mathop{\rm Re}\nolimits}
\def\Im{\mathop{\rm Im}\nolimits}

\def\rme{{\rm e}}
\def\rmi{{\rm i}}

\newcommand{\hc}{{\rm h.c.}}
\newcommand{\bbox}{\lower.2ex\hbox{$\Box$}}
\newsavebox{\uuunit}
\sbox{\uuunit}
    {\setlength{\unitlength}{0.825em}
     \begin{picture}(0.6,0.7)
        \thinlines
        \put(0,0){\line(1,0){0.5}}
        \put(0.15,0){\line(0,1){0.7}}
        \put(0.35,0){\line(0,1){0.8}}
       \multiput(0.3,0.8)(-0.04,-0.02){12}{\rule{0.5pt}{0.5pt}}
     \end {picture}}

\csname @addtoreset\endcsname{equation}{section}
\newcommand{\SU}{\mathop{\rm SU}}

\newcommand{\U}{\mathop{\rm {}U}}

   \usepackage[nosort]{cite}
   \pdfoutput=1
  \usepackage[pdftex]{hyperref}
\usepackage{tcolorbox}

\usepackage{color}
 
\newcommand{\dr}{\raise.3ex\hbox{$\stackrel{\leftarrow}{\delta  }$}{}}
\newcommand{\dl}{\raise.3ex\hbox{$\stackrel{\rightarrow}{\delta }$}{} }
\newcommand{\pl}{\raise.3ex\hbox{$\stackrel{\rightarrow}{\partial }$}{} }

\begin{document}

\begin{titlepage}
\begin{flushright}
CERN-TH-2019-220
\end{flushright}
\vspace{.5cm}
\begin{center}
\baselineskip=16pt
{\LARGE  de Sitter Conjectures in ${\cal N}=1$ Supergravity  
}\\
\vfill
{\large  {\bf Sergio Ferrara}$^{1}$,  {\bf Magnus Tournoy}$^2$ and {\bf Antoine Van Proeyen$^2$}, } \\
\vfill

{\small$^1$ Theoretical Physics Department, CERN CH-1211 Geneva 23, Switzerland\\\smallskip
 INFN - Laboratori Nazionali di Frascati Via Enrico Fermi 40, I-00044 Frascati, Italy\\\smallskip
$^2$   KU Leuven, Institute for Theoretical Physics, Celestijnenlaan 200D, B-3001 Leuven,
Belgium  \\[2mm] }
\end{center}
\vfill
\begin{center}
{\bf Abstract}
\end{center}
{\small
Supergravity theories at $D=4$ allow to formulate the Swampland  de Sitter conjectures
in the complex field space of scalar  components of chiral multiplets.
We formulate the refined de Sitter conjecture by using the K\"{a}hler invariant ${\cal G}$-function
and explore a class of models in the Landscape/Swampland scenario which obey and/or violate such
conjectures. Furthermore we give a new construction of exponential potentials in supergravity. These depend on a chiral superfield with a K\"ahler potential parametrizing an $\SU(1,1)/\U(1)$ geometry. We show that the construction allows for modifications to supergravity theories causing them to obey the de Sitter conjectures.
} \vfill

\hrule width 3.cm
{\footnotesize \noindent e-mails: Sergio.Ferrara@cern.ch, magnus.tournoy@kuleuven.be, antoine.vanproeyen@fys.kuleuven.be }
\end{titlepage}

\addtocounter{page}{1}
 \tableofcontents{}
\newpage
\section{Introduction} \label{sec:Intro}

It is well known that while AdS geometries may provide both broken and unbroken supersymmetry in the vacuum of scalar field configurations,
de Sitter geometry always implies broken supersymmetry in any phase of the theory. The refined  de Sitter conjecture may imply that also
metastable vacua may not exist. In the present note we consider classes of simple theories with broken supersymmetry and see whether
the Swampland criteria are verified.
Our conclusion is that Swampland string conjectures are generically not satisfied in supergravity so that many supergravity theories are not
the Effective Field Theories (EFT) of some ultraviolet complete theory. On the other end these constraints may be avoided if string theory with broken supersymmetry
is studied beyond its weak coupling perturbative regime.

Based on numerous constructions in string theory conjectures have been put forward that pushes (meta-)stable de Sitter vacua into the Swampland, see \cite{Obied:2018sgi,Andriot:2018wzk,Garg:2018reu,Ooguri:2018wrx,Andriot:2018mav} or the review \cite{Palti:2019pca}. We cite from \cite{Ooguri:2018wrx}:
\begin{quote}
 {\bf Refined de Sitter Conjecture.}
{\it A potential $V(\phi)$ for scalar fields in a low energy
effective theory of any consistent quantum gravity must satisfy either,
\begin{equation}
\left|\nabla V\right| \geq \frac{c}{M_p} \cdot V \;,
\label{rdscone}
\end{equation}
or
\begin{equation}
{\rm min} \left( \nabla_i \nabla_j V \right) \leq - \frac{c'}{M_p^2} \cdot V \;,
\label{rdsctwo}
\end{equation}
for some universal constants $c, c' >0$ of order $1$,
 where the left-hand side of (\ref{rdsctwo}) is the minimum eigenvalue of the Hessian $ \nabla_i \nabla_j V$ in an orthonormal frame. }
\end{quote}
The refined conjecture in \eqref{rdscone}, \eqref{rdsctwo} was first formulated in \cite{Garg:2018reu} based on tachyonic tree level de Sitter constructions in IIA/IIB string theory (an earlier suggestion for a modification to \cite{Obied:2018sgi} was made in \cite{Andriot:2018wzk} but was not formulated as a covariant bound on the scalar potential). More general arguments for the refined conjecture that don't rely on specific string theory compactifications were later on given in \cite{Ooguri:2018wrx}.

In this work we reformulate the de Sitter Swampland conjectures into ${\cal N}=1$, $D=4$ supergravity, where the scalars in the conjectures are complex scalars $z^\alpha $. The first conjecture implies that extrema of the potential do not appear for positive cosmological constant:
\begin{equation}
  \nabla _\alpha V = 0 \ \rightarrow   V\leq 0\,.
 \label{nodS}
\end{equation}
There is no precise value for the constant $c$, and thus the above implication is the main part of the conjecture that we can check. For positive potential the full conjecture is that
\begin{equation}
  V>0\ \rightarrow \ \kappa ^{-2} \frac{\partial _\alpha Vg^{\alpha \bar \beta }\partial _{\bar \beta }V}{V^2}\geq \frac12c^2 \,.
 \label{Vg0conjecture1}
\end{equation}
If the first conjecture is not satisfied, the second conjecture may offer an escape for the model. It has thus mainly to be considered when there are vacua with \emph{positive cosmological constant}. They then still \emph{pass the conjectures if the lowest value for second derivatives of the potential is enough negative} (depending on the value of $c'$). Again, since we do not know $c'$, we will check whether there is a negative value.

The conjectures can further be written as constraints on the K\"{a}hler-invariant functional ${\cal G}$, as we will show explicitly below.

The paper contains several short sections made as follows:

In sections \ref{sec:Massunits} and \ref{sec:Massrel} we recall the supergravity mass formulae for non-supersymmetric configurations. In section \ref{sec:Conj} the refined de Sitter
conjectures are formulated in terms of the K\"{a}hler-invariant function ${\cal G}$, the fermionic shift and fermion mass matrices.
In section \ref{sec:Pol} and \ref{sec:onescal} we study the conjectures in the case of the Pol\'{o}nyi model and in the case of R-symmetric potentials for one complex scalar, and obtain bounds on parameters in the theories.
Finally in section \ref{sec:exp} we consider the case of exponential scalar potentials. Models of this kind can describe present stage quintessence. We show how they can be simply constructed by the inclusion of a chiral multiplet with a K\"{a}hler potential typical in $\alpha$-attractor models to the theory.
In particular, we consider the potential induced by nilpotent scalars as in the KKLT construction. These classes of models are more close to stringy potentials and indeed the Swampland conjectures are often satisfied in this case. We further discuss how the construction of single exponential potentials can be used to modify supergravity theories such that they obey the de Sitter conjectures.

\section{Mass units} \label{sec:Massunits}
It is easiest to work first with the engineering dimensions as in \cite[Sec.18.3.1]{Freedman:2012zz}. This uses dimensionless scalars. Explicit $\kappa =M_p^{-1}$ are then introduced in
\begin{equation}
   {\cal K}{:}\ \kappa ^{-2}\,,\qquad g_{\alpha \bar\beta  }{:}\ \kappa ^{-2}\,,\qquad
  W{:}\ \kappa ^{-3}\,.
 \label{kappafactors}
\end{equation}
E.g. for the simple flat K\"{a}hler case we write
\begin{equation}
  {\cal K}= \kappa^{-2} z^\alpha
\delta_{\alpha \bar \beta } \bar z^{\bar \beta }\,.
 \label{Kflat}
\end{equation}
One can redefine at the end $z'
=\kappa^{-1} z$ to avoid factors $\kappa$ in the kinetic term, and $z'$ has then the physical dimension~1.

The potential (here for only chiral multiplets)
\begin{equation}
V=  -3\kappa ^2\rme^{\kappa ^2 {\cal K}}W \overline{W}+
\rme^{\kappa ^2 {\cal K}}\nabla _\alpha  W
    g^{\alpha \bar \beta
}\overline\nabla _{\bar \beta }\overline{W}
 \label{Vwithkappa}
\end{equation}
has then mass dimension 4 as it should be.

The invariant function
\begin{equation}
{\cal G}=\kappa ^2{\cal K} +\log (\kappa ^6W\overline W) \,,
 \label{calGdef}
\end{equation}
is dimensionless, and in terms of this:
\begin{equation}
  V_F= \kappa ^{-4}\rme^{{\cal G}}\left( {\cal G}_\alpha {\cal G}^{\alpha \bar \beta }{\cal G}_{\bar \beta }-3\right) \,,
 \label{VFincalG}
\end{equation}
where ${\cal G}_\alpha =\partial _\alpha {\cal G}$, \ldots  and ${\cal
G}^{\alpha \bar \beta }$ is the inverse of ${\cal G}_{\alpha \bar \beta
}=\kappa ^2g_{\alpha \bar \beta }$.

Example: in this way $\partial _\alpha V g^{\alpha \bar \beta }\partial _{\bar \beta }V$ has, due to the inverse metric, a mass dimension 2 lower than $V^2$. This is thus consistent with (\ref{Vg0conjecture1}). When we go to the $z'$ coordinates this factor $\kappa ^2$ comes from the two derivatives to the scalars.

\section{Mass relations} \label{sec:Massrel}
Several quantities that were introduced to study mass relations in \cite{Ferrara:2016ntj} are useful in this context. E.g. one defines
\begin{equation}
  X\equiv \kappa ^{-2} \frac{\nabla _\alpha W g^{\alpha \bar \beta }\overline\nabla_{\bar \beta  }\overline{W}}{W\overline{W}}\,.
 \label{defX}
\end{equation}
In the ${\cal G}$-formulation, this is
\begin{equation}
  X ={\cal G}_\alpha {\cal G}{}^\alpha  = \kappa ^{-2}{\cal G}_\alpha g^{\alpha \bar \beta }  {\cal G}_{\bar \beta }= {\cal G}_\alpha {\cal G}^{\alpha \bar \beta }  {\cal G}_{\bar \beta }\,,
 \label{defXG}
\end{equation}
where ${\cal G}^{\alpha \bar \beta}$ is the inverse of ${\cal G}_{\alpha \bar \beta}= \kappa ^2 g_{\alpha \bar \beta}$ and is thus dimensionless. Similarly, we define here
\begin{equation}
  {\cal G}^\alpha \equiv  {\cal G}^{\alpha \bar \beta }  {\cal G}_{\bar \beta }= \kappa ^{-2}g^{\alpha \bar \beta }  {\cal G}_{\bar \beta }\,.
 \label{Gup}
\end{equation}
This allows to write the potential as
\begin{equation}
  V = \kappa ^{-4}\rme^{{\cal G}}(X-3)\,.
 \label{VGX}
\end{equation}
The full holomorphic mass matrix for the fermions is
\begin{equation}
 M_{\alpha \beta }=  \sqrt{\frac{W}{\overline{W}}}
 \rme^{{\cal G}/2}\left[{\cal G}_{\alpha \beta }+ \frac{X-2}{X}{\cal G}_\alpha {\cal G}_\beta\right]\,,\qquad{\cal G}_{\alpha \beta }=\nabla _\alpha \partial _\beta {\cal G} \,.
 \label{fullM}
\end{equation}
Since the overall factors are going to cancel at the end, we define
\begin{equation}
  M_{\alpha \beta }=  \sqrt{\frac{W}{\overline{W}}}= \sqrt{\frac{W}{\overline{W}}}\rme^{{\cal G}/2}{\cal M}_{\alpha \beta }\,,\qquad {\cal M}_{\alpha \beta }={\cal G}_{\alpha \beta }+ \frac{X-2}{X}{\cal G}_\alpha {\cal G}_\beta\,.
 \label{MtocalM}
\end{equation}
It appears in the derivative of the potential
\begin{align}
 V_\alpha \equiv  \partial _\alpha V =&\kappa ^{-4}\rme^{{\cal G}}\left[{\cal G}_{\alpha \beta }{\cal G}^\beta +(X-2){\cal G}_\alpha \right]=\kappa ^{-4}\rme^{{\cal G}} {\cal M}_{\alpha \beta }{\cal G}^\beta  \,.
\label{dalphaV}
\end{align}
For analyzing the condition (\ref{rdsctwo}), we provide the second derivatives
\begin{align}
\kappa ^4V_{\alpha\bar \alpha }=&\rme^{{\cal G}}\left[g_{\alpha \bar \alpha }(X-2)-{\cal G}_\alpha {\cal G}_{\bar \alpha }+\left({\cal G}_{\alpha \beta }+{\cal G}_\alpha {\cal G}_\beta \right)g^{\beta \bar \beta }\left( {\cal G}_{\bar \alpha \bar \beta }+{\cal G}_{\bar \alpha }{\cal G}_{\bar \beta }\right) + R_{\alpha \bar \alpha} {}^{\beta\bar  \beta } {\cal G} _\beta  \,{\cal G}_{\bar \beta}\right]\,,\nonumber\\
\kappa ^4V_{\alpha \beta }=&\rme^{{\cal G}}\left[-\left({\cal G}_{\alpha \beta }+{\cal G}_\alpha {\cal G}_\beta \right) +\left( {\cal G}_{\alpha \beta \gamma }+{\cal G}_\alpha {\cal G}_\beta {\cal G}_\gamma +3{\cal G}_{(\alpha \beta }{\cal G}_{\gamma )}\right) {\cal G}^\gamma\right] \,.
\label{Vderiv}
\end{align}
\section{Conjectures in \texorpdfstring{${\cal N}=1$ }{N=1} supergravity} \label{sec:Conj}
Using (\ref{VGX}) and  (\ref{dalphaV}), the left-hand side of (\ref{Vg0conjecture1}) can be written as
\begin{align}
 \kappa ^{-2} \frac{\partial _\alpha Vg^{\alpha \bar \beta }\partial _{\bar \beta }V}{V^2}= & \frac{ {\cal M}_{\alpha \beta }{\cal G}^\beta  {\cal G}^{\alpha \bar \alpha } {\cal M}_{\bar \alpha\bar  \beta }{{\cal G}}^{\bar \beta }}{(X-3)^2} \,,
\label{conjectureN1frac}
\end{align}
 and thus the first of the de Sitter Swampland conjectures can be formulated as
\begin{align}
\sqrt{2{\cal M}_{\alpha \beta }{\cal G}^\beta  {\cal G}^{\alpha \bar \alpha } {\cal M}_{\bar \alpha\bar  \beta }{{\cal G}}^{\bar \beta }}\geq  c (X-3) \,.
\label{conjectureN1}
\end{align}
An alternative expression using
\begin{align}
{\cal H}\equiv \log (\kappa ^4V) & ={\cal G}+\log(X-3)\,,\qquad {\cal H}_{\alpha}=\partial _\alpha\log (\kappa ^4V)=  {\cal G}_\alpha+\frac{\partial _\alpha X}{X-3}\,,
\label{logVd}
\end{align}
is
\begin{align}
 \kappa ^{-2} \frac{\partial _\alpha Vg^{\alpha \bar \beta }\partial _{\bar \beta }V}{V^2}= {\cal H}_\alpha {\cal G}^{\alpha\bar\beta} {\cal H}_{\bar\beta}= X + \frac{1}{X-3}\left({\cal G}{}^\alpha \partial _\alpha X+\hc\right)+ \frac{1}{(X-3)^2}\partial _\alpha X{\cal G}^{\alpha \bar \alpha }\partial _{\bar \alpha }X \,.
\label{alternativeconj}
\end{align}
Thus for $V>0$, according to the first conjecture we should have ${\cal H}_\alpha {\cal H}^\alpha \geq \ft12c^2 $.

Note that, using $\nabla _\alpha {\cal G}^\beta  =\delta _\alpha ^\beta $:
\begin{equation}
  \partial _\alpha X = {\cal G}_{\alpha \beta }{\cal G}{}^\beta +{\cal G}_\alpha = {\cal M}_{\alpha \beta }{\cal G}{}^\beta +(3-X){\cal G}_\alpha\,,
 \label{dalphaX}
\end{equation}
and therefore the connection with the formulation in \eqref{conjectureN1frac} can be readily made by
\begin{equation}
{\cal H}_\alpha=\frac{\cal M_{\alpha\bar\beta}{\cal G}{}^\beta}{X-3}\,.
\label{HandM}
\end{equation}
For the second conjecture we introduce latin indices $a,b$ to denote the holomorphic as well as anti-holomorphic coordinates. Since $\partial _a V= {\cal H}_a V$, we find for the second derivatives of the potential
\begin{align}
\nabla_a\partial_{b}V&={\cal H}_{ab} V+{\cal H}_a{\cal H}_{b}V\,,
\label{secderV}
\end{align}
where ${\cal H}_{ab}=\nabla_a \partial_b {\cal H}$.
The second de Sitter conjecture for the canonically normalized scalars becomes
\begin{equation}
{\rm min}\left( \frac{\partial z^a}{\partial \phi^i}\frac{\partial z^b}{\partial\phi^j}\left[{\cal H}_{ab}+{\cal H}_{a}{\cal H}_b\right]\right) \leq - \kappa^2 c' \,.
\label{conjectureN2}
\end{equation}
In terms of ${\cal G}$:
\begin{align}
\nabla_a\partial_{b}V=&V\left({\cal G}_{ab}+{\cal G}_a {\cal G}_{b}+\frac{2}{X-3}{\cal G}_{(a}\partial_{b)}X+\frac1{X-3}\nabla_a\partial_{b} X\right)\,,\nonumber\\
\partial_b X=&{\cal G}^a{\cal G}_{ba}={\cal G}^{\bar\alpha}{\cal G}_{b\bar\alpha}+{\cal G}^\alpha {\cal G}_{b\alpha}\,,\nonumber\\
\nabla_a \partial_b X=&{\cal G}_{a}{}^c{\cal G}_{bc}+{\cal G}^c{\cal G}_{abc}={\cal G}_{a}{}^\alpha{\cal G}_{b\alpha}+{\cal G}_{a}{}^{\bar\alpha}{\cal G}_{b\bar\alpha}+{\cal G}^\alpha{\cal G}_{ab\alpha}+{\cal G}^{\bar\alpha}{\cal G}_{ab\bar\alpha}
\,.
\label{secderX}
\end{align}

\section{Pol\'onyi model} \label{sec:Pol}
The Pol\'onyi model \cite{Polonyi:1977pj} is of importance for the description of supersymmetry breaking within the supergravity framework. It contains a single chiral multiplet on a flat K\"{a}hler manifold guided by the superpotential $W=\kappa^{-3}\mu\left(z+\beta\right)$. Therefore
\begin{equation}
{\cal G}=z\bar z+\log(\mu^2|z+\beta|^2),\quad {\cal G}_z=\bar z+\frac1{z+\beta},\quad {\cal G}_{z\bar z}=1,\quad {\cal G}_{zz}=-\frac1{(z+\beta)^2}\,.
\label{PolG}
\end{equation}
The parameters $\mu,\, \beta$ are taken to be real. Since $\beta $ is of order $1$, we will restrict to $0\leq\beta<2$.

We write $z$ in terms of two real canonically normalized fields $z=\frac{1}{\sqrt2}\left(x+\rmi y\right)$. In the direction $y$ the minimum appears for $y=0$. The potential $V$ is then solely a function of $x$. We find
\begin{align}
X&=\left|{\cal G}_z\right|^2=\left(\frac{x}{\sqrt{2}}+\frac{1}{\frac{x}{\sqrt{2}}+\beta}\right)^2\,,\nonumber\\
V(x)&=\kappa^{-4}\mu ^2 \rme^{x^2/2} \left(\frac{x}{\sqrt{2}}+\beta\right)^2 \left(\left(\frac{x}{\sqrt{2}}+\frac{1}{\frac{x}{\sqrt{2}}+\beta}\right)^2-3\right)\,,\nonumber\\
{\cal M}_{zz}&=-\frac{1}{(\frac{x}{\sqrt2}+\beta)^2}+(X-2)\,,\nonumber\\
{\cal M}_{zz}{\cal G}^z&=\frac{\left(x^2+\sqrt{2} \beta  x+2\right) \left( x^3+\sqrt2 \beta  x^2-4\sqrt2 \beta\right)}{2\sqrt2 \left( x+\sqrt2\beta\right)^2}\,.
\label{PolXM}
\end{align}
Since the potential is proportional to $\mu^2 $, this parameter is irrelevant in (\ref{Vg0conjecture1}).

\begin{figure}[th]
    \centering
    \includegraphics[width=0.75\textwidth]{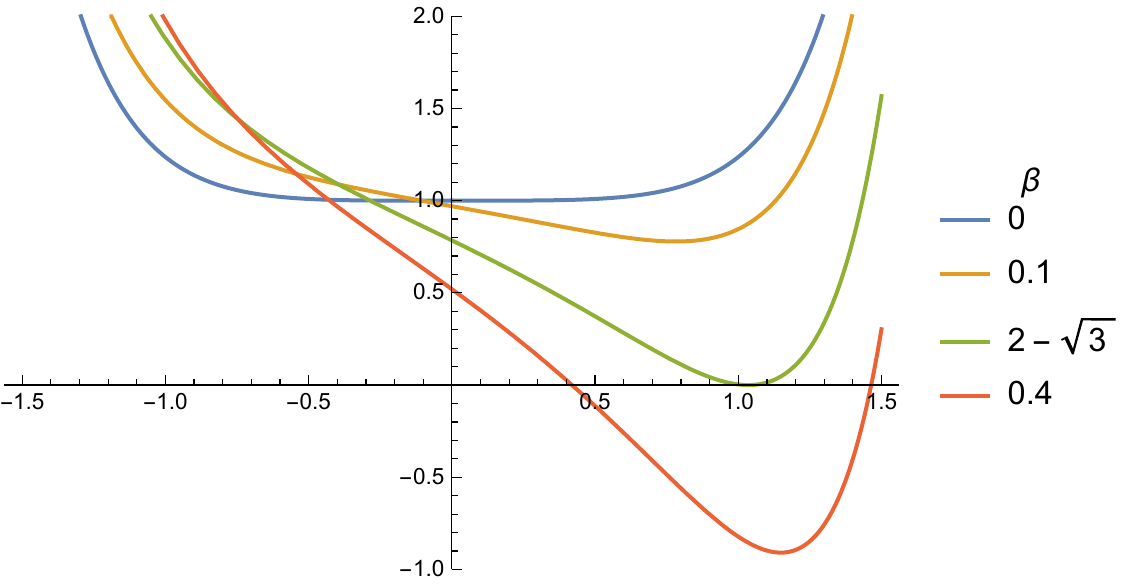}
    \caption{The scalar potential of the Pol\'onyi model for $\beta=\left\{0,\, 0.1,\, 2-\sqrt3,\, 0.4\right\}$. For $0\leq \beta\leq 2-\sqrt3$ the minimum is positioned at $V\geq 0$. When $\beta>2-\sqrt3$ there is a minimum at $V<0$.}
    \label{potpol}
\end{figure}

\begin{figure}[bh]
    \centering
    \includegraphics[width=0.5\textwidth]{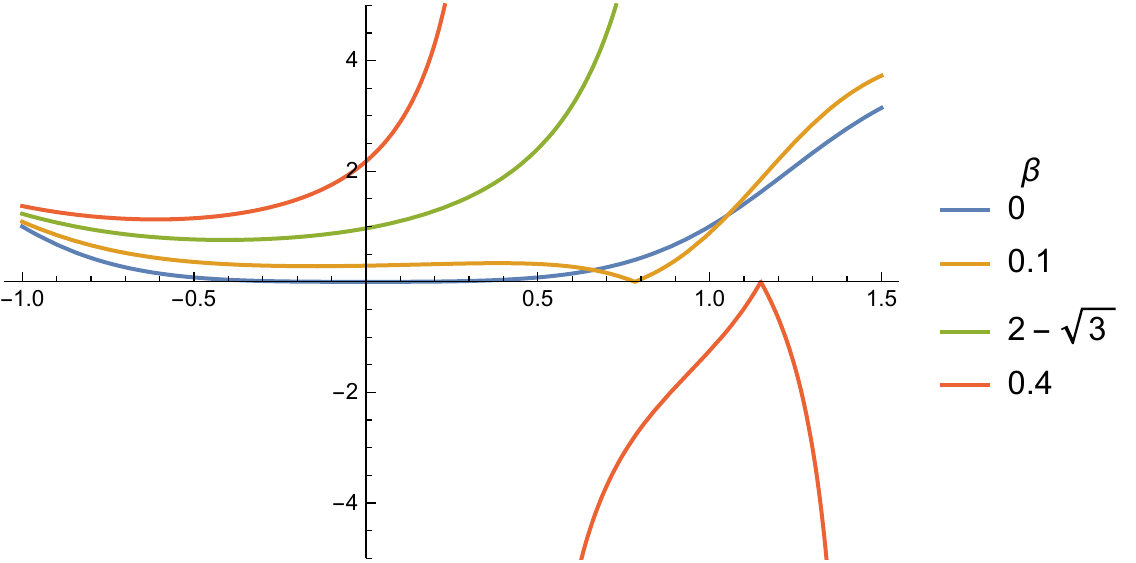}
    \caption{The function $|\nabla V|/V$ for the Pol\'onyi model}
    \label{plotpoldsconj}
\end{figure}

The formula in \eqref{conjectureN1frac} then becomes
\begin{equation}
\kappa^{-1}\frac{|\nabla V|}{V}=\frac{| x^5+2\sqrt2 \beta  x^4+2  \left(\beta ^2+1\right) x^3-2\sqrt2 \beta  x^2-8 \beta ^2 x-8\sqrt2 \beta |}{2\sqrt2 \left(x^4+2 \sqrt{2} \beta  x^3+2 \left(\beta ^2-1\right) x^2-8 \sqrt{2} \beta  x+4-12 \beta ^2\right)}.
\label{dSPol}
\end{equation}
The real root of \eqref{dSPol} for $0\leq \beta <2$ is
\begin{equation}
x= \frac{\sqrt2}{3} \left[-\beta+3\beta ^{1/3}\left(1-\frac{\beta ^2}{27}+\sqrt{1-2 \frac{\beta ^2}{27}} \right)^{1/3}+\frac{\beta ^{5/3}}{3 \left(1-\frac{\beta ^2}{27}+\sqrt{1-2 \frac{\beta ^2}{27}} \right)^{1/3}}\right]\,.
\label{Polroots}
\end{equation}
One can check that the single extremum for $0\leq\beta<2$ is a minimum. By looking at the zeros of the potential for $0\leq \beta < 2$
\begin{equation}
V(x_0) = 0\ : \ x_0=\frac1{\sqrt2}\left(\sqrt3-\beta\pm\sqrt{-1+2\sqrt3\beta+\beta^2}\right)
\label{Polpotroots}
\end{equation}
we see the lowest value of $\beta$ for which real zero value for the potential exists is $\beta=2-\sqrt3$. In that case $x_0=\sqrt2\left(\sqrt3-1\right)$ and, as one can show, it is then also the minimum of the potential. In fact, it is the value chosen by Pol\'onyi, who was looking for a Minkowski vacuum. This model has also been studied in \cite{Kallosh:2002gf}, where Figure \ref{potpol} is also drawn.
One can see in that Figure (where $\kappa =\mu =1$)
that the potential is negative in the interval between the zero modes in \eqref{Polpotroots}, where it will reach its minimum value. Thus we find
\begin{align}
  0\leq \beta <2-\sqrt3= 0.27\ :\ & \mbox{minimum at }V>0\ :\  (\ref{nodS})\ \mbox{violated}\,, \nonumber\\
  \beta \leq 2-\sqrt3 \ :\ & \mbox{minimum at }V\leq 0\ :\  (\ref{nodS})\ \mbox{satisfied}\,,
\end{align}
The value of the ratio $|\nabla V|/V$ is shown in Figure \ref{plotpoldsconj}.

There is only one extremum which is a minimum and the second derivative is everywhere positive. Thus for $0\leq \beta<2-\sqrt3$ both the de Sitter conjectures \eqref{rdscone},\eqref{rdsctwo} are violated.
\newpage
\section{R-symmetric one scalar model} \label{sec:onescal}
We consider a flat K\"{a}hler model with one scalar and $R$-symmetry. Then the superpotential should be homogeneous, i.e. $W= \kappa ^{-3} z^\alpha $. The case $\alpha =1$ thus overlaps with $\beta =0$ in the previous section. We thus have
\begin{align}
  {\cal G}= & z\bar z + \alpha \log(z\bar z)\,,\qquad {\cal G}_z = \frac{1}{z}(z\bar z+\alpha )\,,\qquad {\cal G}_{z\bar z }=1\,,\qquad  {\cal G}_{zz}= -\frac{\alpha }{z^2}\,.
\label{modelG}
\end{align}
In this case, nearly everything depends only on
\begin{equation}
   \rho =z\bar z\,.
 \label{defrho}
\end{equation}
This is in fact the statement of $R$-symmetry.
We find
\begin{align}
  X= &= |{\cal G}_z |^2= \rho ^{-1}\left(\alpha ^2 + 2\alpha \rho +\rho ^2\right)\,, \nonumber\\
  V= & \kappa ^{-4}\rme^\rho \rho ^{\alpha -1}\left(\alpha ^2 + (2\alpha-3) \rho +\rho ^2\right)\,,\nonumber\\
  {\cal M}_{zz}=&-\frac{\alpha }{z^2}+(X-2)\frac{{\cal G}_z}{\overline{{\cal G}}_{\bar z}}\,.
\label{firstrelationstoy}
\end{align}
We thus find that $V$ is everywhere positive if $\alpha >3/4$.

\begin{figure}[ht]
    \centering
    \includegraphics[width=0.75\textwidth]{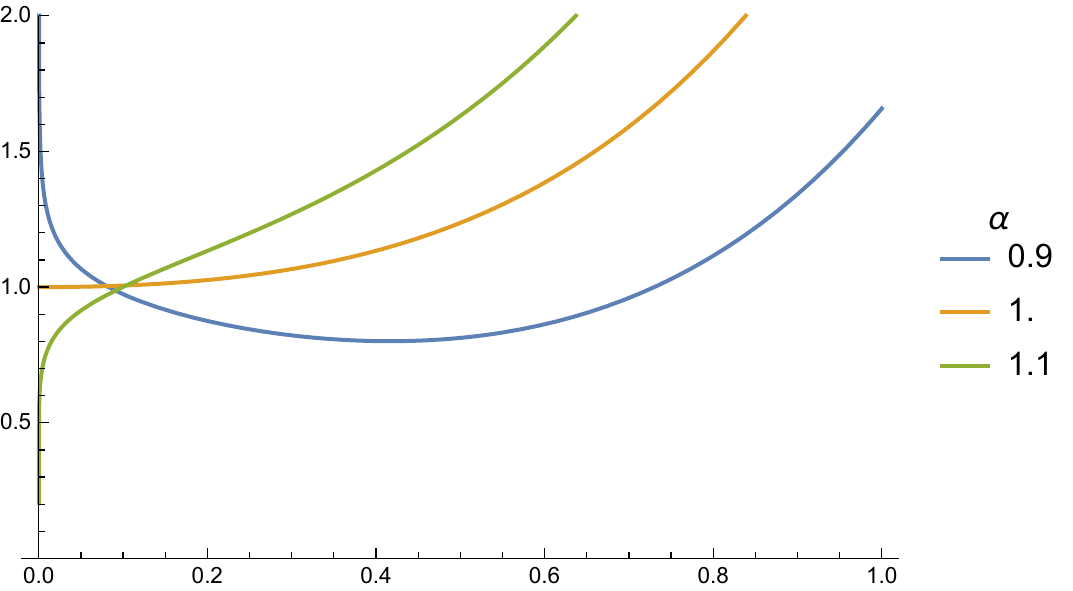}
    \caption{The scalar potential $V(\rho)$ of the one scalar model for values $\alpha\sim 1$}
    \label{pot1}
\end{figure}

Since ${\cal G}_z/{\overline{{\cal G}}_{\bar z}}= \bar z/z = z^{-2}\rho $, we find
\begin{align}
  {\cal M}_{zz} = & \frac{1}{z^2}\left[\alpha (\alpha -1)+ 2 (\alpha -1)\rho +\rho ^2\right]\,,\nonumber\\
 {\cal M}_{zz}\overline{{\cal G}}{}^z=  & \frac{1}{\rho z}\left[\alpha^2 (\alpha -1)+3\alpha (\alpha -1)\rho +(3\alpha -2)\rho ^2+\rho ^3\right]\,.
\label{calMtoy}
\end{align}
Since the metric is trivial and the right-hand side of  (\ref{conjectureN1frac}) is then a modulus squared, we can write
\begin{align}
  \kappa ^{-1}\frac{|\nabla_z V|}{V} & =|\rho ^{-1/2}|\frac{|\alpha^2 (\alpha -1)+3\alpha (\alpha -1)\rho +(3\alpha -2)\rho ^2+\rho ^3|}{\alpha ^2 + (2\alpha-3) \rho +\rho ^2}\,.
\label{relationtoy}
\end{align}
The overall factor $\rho^{1/2}$ appears from the change of variables from $z$ to $\rho $.
\begin{equation}
\kappa^{-1} |\nabla_z V|=\kappa^{-1}|\rho^{1/2}\nabla_\rho V|=\kappa^{-4}|\rme^\rho\rho ^{\alpha-3/2}||\alpha^2 (\alpha -1)+3\alpha (\alpha -1)\rho +(3\alpha -2)\rho ^2+\rho ^3|\,.
\label{changevar}
\end{equation}
The first derivative of the potential has 4 roots
\begin{equation}
|\nabla V|=0 \Leftrightarrow \rho \to \left\{0,1-\alpha -\sqrt{1-\alpha },1-\alpha +\sqrt{1-\alpha },-\alpha\right\}\,.
\label{rootsV}
\end{equation}
Whenever the first derivative of the potential has a root the first de Sitter conjecture can be trivially violated. It is therefore interesting to study these specific points. The overall factor $\rho^{\alpha-3/2}$ in \eqref{changevar} causes $\rho=0$ to be a root when $\alpha>3/2$. But in the de Sitter conjecture the overall factor is reduced to $\rho^{-1/2}$, so there is no effect of this root on the violation of the first de Sitter conjecture. The second and third root are only real for $\alpha\leq 1$. There is thus a distinct behaviour around the value $\alpha=1$, which can also be seen from Figure \ref{pot1}.
\subsection{Evaluation of first conjecture}
\begin{figure}[H]
    \centering
    \includegraphics[width=0.75\textwidth]{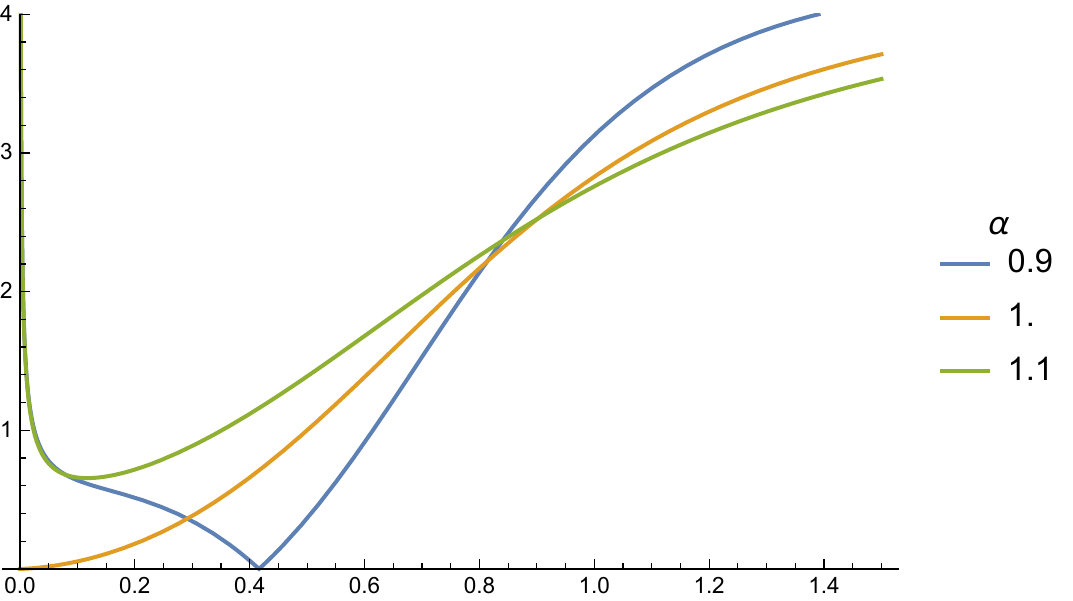}
    \caption{The function $|\nabla V|/V$ of the one scalar model for values $\alpha\sim 1$}
    \label{dSconj1_1}
\end{figure}

\emph{For $\alpha >1$} there is no vacuum, and thus the \emph{first condition can be satisfied.} To consider the condition once a value of $c$ is imposed, see Figure \ref{dSconj1_1}. The minimum of $|\nabla V|/V\sim {\cal O}(1)$.

For $\alpha \leq 1$, the third root in (\ref{rootsV}) occurs for a positive $\rho $ (or $\rho =0$ for $\alpha =1$) and thus defines an extremum. As mentioned above the potential is positive for $\alpha >3/4$, and thus we can identify different classes of potentials.

\emph{For $\ft34<\alpha\leq 1$}, the potential is positive for  $\alpha >3/4$ and thus we have an extremum with a positive $V$, hence a \emph{violation} of  (\ref{nodS}).

\begin{figure}[ht]
    \centering
    \includegraphics[width=0.75\textwidth]{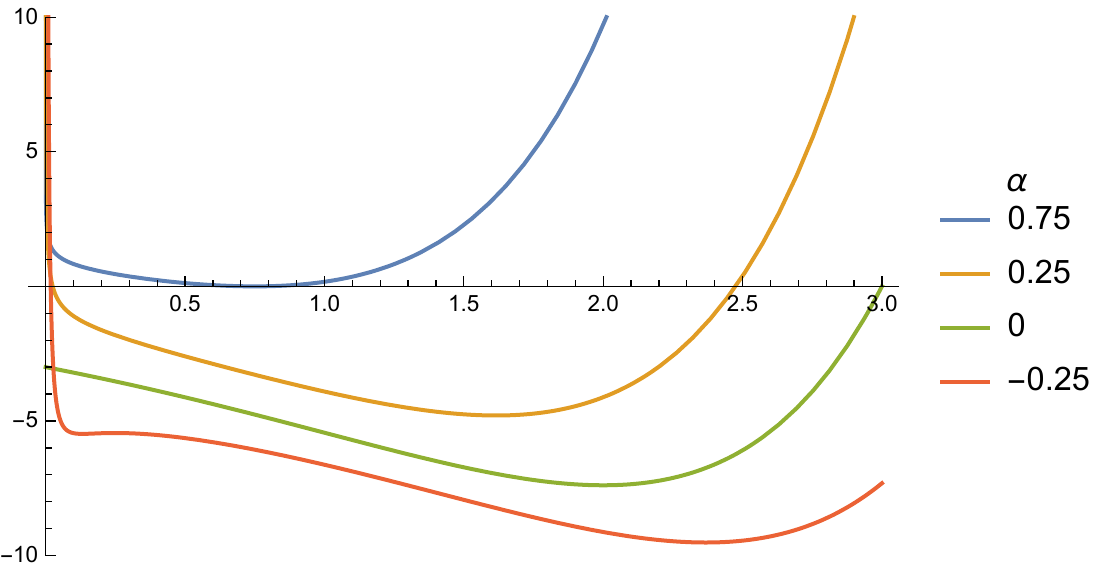}
    \caption{The scalar potential $V(\rho)$ of the one scalar model for values $\alpha \leq 3/4$}
    \label{pot2}
\end{figure}

For $\alpha \leq 3/4$ the potential becomes negative in the following range of values for $\rho$
\begin{equation}
V(\rho)\leq 0\quad \text{for}\quad \frac12\left(3-2\alpha-\sqrt{3}\sqrt{3-4\alpha}\right)\leq \rho\leq \frac12\left(3-2\alpha+\sqrt{3}\sqrt{3-4\alpha}\right)\,. \label{negint}
\end{equation}
Different cases are shown in Figure \ref{pot2}.

\begin{figure}[ht]
    \centering
    \includegraphics[width=0.75\textwidth]{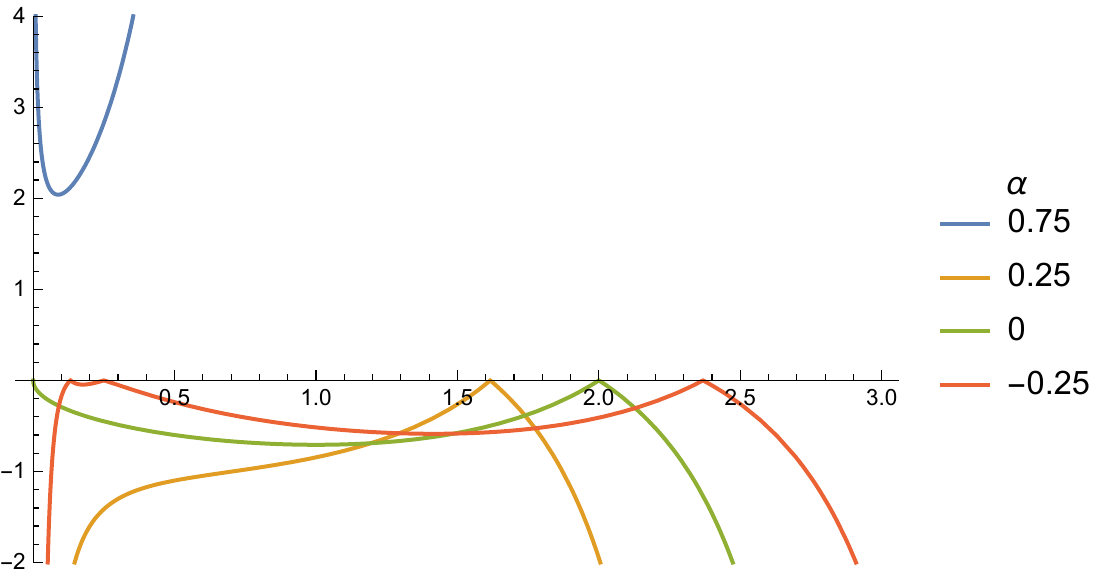}
    \caption{The function $|\nabla V|/V$ of the one scalar model for values $\alpha\leq 3/4$}
    \label{dSconj1_2}
\end{figure}

\emph{For $0\leq\alpha\leq \ft34 $} the potential has only one extremum positioned at $\rho=1-\alpha+\sqrt{1-\alpha}$ which falls in the interval of negative potential \eqref{negint}. We conclude that the first de Sitter conjecture \emph{is not violated by the extrema of these potentials.} The ratio is shown in Figure \ref{dSconj1_2}.

\emph{For $\alpha<0$} the potential has three extrema positioned at
\begin{equation}
\rho=\left\{-\alpha,1-\alpha-\sqrt{1-\alpha},1-\alpha+\sqrt{1-\alpha}\right\}\,.
 \label{extremarho}
\end{equation}
One can check that for $\alpha\leq 0$
\begin{align}
\kappa^4V(-\alpha)&=-3 \rme^{-\alpha}(-\alpha)^{\alpha} \leq 0\,,\nonumber\\
\kappa^4V(1-\alpha-\sqrt{1-\alpha})&=\rme^{-\alpha -\sqrt{1-\alpha }+1} \left(-\alpha -\sqrt{1-\alpha }+1\right)^{\alpha -1} \left(2 \alpha +\sqrt{1-\alpha }-1\right)\leq 0\,,\nonumber\\
\kappa^4V(1-\alpha+\sqrt{1-\alpha})&=\rme^{-\alpha +\sqrt{1-\alpha }+1}  \left(-\alpha +\sqrt{1-\alpha }+1\right)^{\alpha -1}\left(2 \alpha -\sqrt{1-\alpha }-1\right)\leq 0\,.
\label{extpot}
\end{align}
So again we see that around the extrema the potentials are negative which implies that the \emph{first de Sitter conjecture can be satisfied} when $\alpha\leq 0$.

Thus, we found only a problem with the conjecture for the range $\ft34<\alpha\leq 1$.
\subsection{Evaluation of second conjecture}

To evaluate the second conjecture \eqref{rdsctwo} we look for the lowest eigenvalue of the matrix $\partial_i\partial_j V$ where the indices run over the canonically normalized fields $x=\sqrt2\Re(z)$, $y=\sqrt2\Im(z)$. It is easiest to first consider the second derivative of the potential to the fields $\sqrt\rho$ and $\varphi$ : $z=\sqrt{\rho}\rme^{\rmi\varphi}$. Since the potential does not depend on $\varphi$ there is only one of the second derivatives that is non-zero
\begin{align}
V''\equiv\left(\frac{\partial}{\partial\sqrt\rho}\right)^2V&=2\, \rme^\rho \rho^{-2+\alpha}\Big(\alpha^2\left(3-5\alpha+2\alpha^2\right)+\alpha\rho\left(3-11\alpha+8\alpha^2\right)\nonumber\\
&\quad+\rho^2\left(-2-7\alpha+12\alpha^2\right)+\rho^3\left(-1+8\alpha\right)+2\rho^4\Big)\,.
\label{secderpot}
\end{align}
Considering polar coordinates $(\sqrt\rho, \,\varphi )$, there is also a second covariant derivative $V_{\varphi \varphi }$ proportional to $V'$. The eigenvalues of the matrix of covariant derivatives with respect to $x$ and $y$ are then
\begin{equation}
\lambda=\frac12\left\{ \rho ^{-1/2}V' ,\, V''\right\}\,.
\label{eignv}
\end{equation}
where the factor $1/2$ originates from the factors $\sqrt{2}$ in the definition of $x$ and $y$ (in order that they have canonical kinetic energy).

\begin{figure}[htb]
    \centering
    \includegraphics[width=0.75\textwidth]{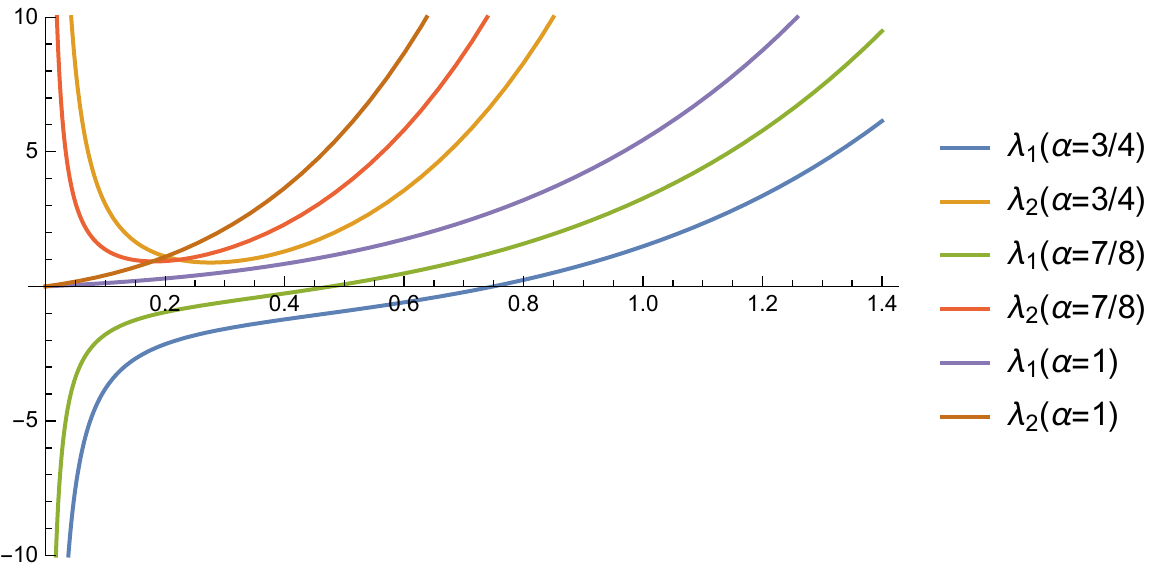}
    \caption{The eigenvalues $\left\{\lambda_1,\,\lambda_2\right\}=\frac12\left\{ \rho ^{-1/2}V' ,\, V''\right\}$ of the one scalar model for $\alpha=7/8$. At the extremum none of the eigenvalues are negative $\left\{0,\,\lambda_2\right\}\geq 0$.}
    \label{dSconj2}
\end{figure}

At the extremum, the non-zero eigenvalue is thus proportional to $V''$. In Figure~\ref{dSconj2}, we draw the eigenvalues for different values of $\alpha$. To satisfy the second conjecture \eqref{rdsctwo}, the smallest one should be negative, which is not satisfied e.g. in the extremum where $\lambda _1=0$, and $\lambda _2>0$.

We conclude that the second conjecture does not  lead to an escape for the troublesome range $\ft34<\alpha\leq 1$.

\newpage

\section{Models with exponential scalar potentials} \label{sec:exp}
One class of models that can trivially satisfy the conjecture in \eqref{rdscone} are the single exponential scalar potentials
\begin{equation}
V(\phi)=V_0\,\rme^{\xi\phi}\,,\qquad V_0>0\,.
\label{exppot}
\end{equation}
These models can describe the effects of a slowly changing vacuum energy caused by a rolling field. They are therefore the most simplest of the present stage quintessence models and are in agreement with observations when $\xi\lesssim 0.6$ \cite{Agrawal:2018own}. In this section we show how such models and extensions thereof can be constructed in a supergravity setup. We hereby use and expand on the work done in \cite{Brax:1999gp,Copeland:2000vh,Chiang:2018jdg}.

We start with K\"{a}hler and superpotential, $K$ and $W$, describing the dynamics of a set of chiral superfields $z^\alpha$ by a scalar potential $V$. The next step is to add a chiral superfield $\Phi$ to the K\"{a}hler potential of the previous theory
\begin{equation}
\widehat{K}=K-a\kappa ^{-2}\log(\Phi+\bar\Phi)\,,\quad a>0.
\label{KPhi}
\end{equation}
The K\"{a}hler potential of the field $\Phi$ in \eqref{KPhi} parametrizes an $\SU(1,1)/\U(1)$ symmetric geometry and is well-known in the literature where it is used in inflationary $\alpha$-attractor models \cite{Kallosh:2013yoa} with $a=3\alpha $. The chiral superfield $\Phi$ appears in the superpotential with a power law
\begin{equation}
\widehat{W}=\Phi^{-b/2}W\,.
\label{WPhi}
\end{equation}
The scalar potential of the extended theory is
\begin{equation}
\widehat{V}=\left(\Phi+\bar\Phi\right)^{-a}|\Phi|^{-b}\left(V+\frac{\rme^{\kappa ^2K}}{a\kappa^4|\Phi|^2}\left[\frac{b}{2}\left(\frac{b}{2}+a\right)\left(\Phi+\bar\Phi\right)^2+a^2|\Phi|^2\right]\right)\,,
\label{VPhi}
\end{equation}
where we used $V$ to describe the scalar potential of the original theory. The potential in \eqref{VPhi} is minimal in the $\Phi $ plane for $\Im \Phi = 0$. Considering therefore only the real part of the scalar field, one finds that after canonically normalizing
\begin{equation}
\Phi+\bar\Phi=\rme^{-\sqrt{2/a}\phi}\,,
\label{cannorm}
\end{equation}
the scalar potential takes the form\footnote{Since the kinetic term
\[
\frac{a\kappa ^{-2}}{(\Phi+\bar\Phi)^2}\partial\Phi\partial\bar\Phi
\]
is scale invariant. There is a freedom in redefining the field which allows us to fix the scale of the potential.}
\begin{equation}
\widehat{V}=\rme^{\sqrt{2\gamma}\phi}V+\gamma \kappa^{-2}|m_{3/2}|^2\,,
\label{fullpot}
\end{equation}
with $\gamma=(a+b)^2/a$ and the gravitino mass given by
\begin{equation}
|m_{3/2}|^2=\kappa^{-2}\rme^{\widehat{\cal G}}=\kappa^4\rme^{\sqrt{2\gamma}\phi+\kappa ^2 K}|W|^2\,.
\label{gravitmass}
\end{equation}
By demanding that there exists a stable minimum for the fields $z^\alpha,\bar z^{\bar\alpha}$, we find the condition
\begin{equation}
\partial_\alpha \widehat{V}=0\quad\Leftrightarrow\quad \partial_\alpha V+\gamma\kappa^2\partial_\alpha(\rme^{\kappa ^2K} |W|^2)=0\,,
\label{cond1}
\end{equation}
and its holomorphic counterpart. Using the expression of the scalar potential and $F^\alpha=\rme^{\kappa ^2K/2}g^{\alpha\bar\beta}\nabla_{\bar\beta} \bar W$, the condition in \eqref{cond1} can be rewritten as
\begin{equation}
\gamma F^\beta \nabla_\alpha \bar{F}_\beta + (2-\gamma )V_\alpha =0\,.
\label{cond2}
\end{equation}
Whenever the condition in \eqref{cond2} can be satisfied, at that specific point the only dynamical field will be the rolling scalar following the exponential potential
\begin{equation}
\widehat{V}=\rme^{\sqrt{2\gamma}\phi}\left(V_0+\gamma\kappa^2 \rme^{\kappa ^2K}|W|^2|_{z,\bar z=z_0,\bar z_0}\right)\,.
\label{newV}
\end{equation}
Notice from \eqref{cond2} that when the vacuum of the scalar potential $V$ is supersymmetric ($F^\alpha =0$) the condition (\ref{cond1}) is immediately satisfied. Therefore in the case of supersymmetric vacua the addition of the chiral multiplet $\Phi$ leaves the position of the vacua unperturbed. Furthermore notice that for supersymmetric vacua, the uplift to a de Sitter vacuum can only happen when $\gamma>3$. However in this case the phenomenological constraint $\xi \lesssim 0.6$ is not satisfied. This implies that in order for the models just described to be phenomenologically viable the breaking of supersymmetry, needed for the de Sitter uplift, cannot be solely caused by the quintessence field $\Phi$.

\vspace{3mm}
The inclusion of a quintessence scalar, in the way we presented in (\ref{KPhi}),~(\ref{WPhi}), provides a simple way to escape the violation of the de Sitter conjectures. The effects on the original model are minimal as can be seen from \eqref{fullpot} which can also be written as
\begin{equation}
\widehat{V}=e^{\sqrt{2\gamma}\phi+\kappa^2 K}\left(\nabla_{\alpha}W g^{\alpha\bar\beta}\nabla_{\bar\beta}\overline W-(3-\gamma)\kappa^2|W|^2\right)\,.
\label{extexp}
\end{equation}
Therefore when $\gamma$ is sufficiently small the main features of the theory will be preserved. From \eqref{rdscone} one can see that the lower bound on $\gamma$ in order for the first de Sitter conjecture to be everywhere satisfied is
\begin{equation}
\gamma\geq\left(\frac{c^2}{2}-\frac{\partial _\alpha \widehat V g^{\alpha \bar \beta }\partial _{\bar \beta }\widehat V}{\kappa ^2\widehat V^2}\right)\quad \text{for }\widehat{V} > 0\,.
\label{constrgammma}
\end{equation}
\vspace{3mm}

\textbf{Example 1.} The simplest example is the case where the uplift is caused by a nilpotent superfield $X^2=0$. Since this multiplet does not contain any scalars, condition \eqref{cond2} is trivially satisfied. The K\"{a}hler and superpotential are given by
\begin{equation}
\widehat{K}=X\bar X-a\kappa ^{-2}\log(\Phi+\bar\Phi)\,,\quad W=\Phi^{-b/2}(m\kappa ^{-2}+fX)\,.
\label{KWex}
\end{equation}
The scalar potential of this model is
\begin{equation}
\widehat{V}=\rme^{\sqrt{2\gamma}\phi}\left(f^2-\kappa ^{-2}m^2(3-\gamma)\right)\,.
\label{NilpV}
\end{equation}
The scalar potential is similar to the one found in \cite{Dudas:2015eha,Bergshoeff:2015tra,Hasegawa:2015bza}. One important difference is that for the model in \eqref{NilpV} the difference between the supersymmetry breaking scale and the gravitino mass is decoupled from the vacuum energy by the presence of an extra parameter $\gamma$ associated to the quintessence.
\vspace{3mm}

\textbf{Example 2.} Looking back at the Pol\'onyi model in Figure \ref{sec:Pol}, we can see how the addition of a quintessence scalar changes the picture. The K\"{a}hler-invariant functional ${\cal G}$ is
\begin{equation}
{\cal G}=z\bar z-a\log(\Phi+\bar\Phi)+\log(\mu^2|z+\beta|^2)+\frac{b^2}{4}\log(\Phi\bar\Phi)\,.
\label{PolQ}
\end{equation}
After going to canonically normalized coordinates $z=\frac1{\sqrt2}\left(x+\rmi y\right)$, $\Phi+\bar\Phi=\rme^{-\sqrt{2/a}\phi}$ one can check that $y=0$ forms a minimum, and the scalar potential becomes
\begin{equation}
\widehat{V}(x,\phi )=\kappa^{-4}\mu ^2 \rme^{\sqrt{2\gamma}\phi} \rme^{x^2/2} \left(\frac{x}{\sqrt{2}}+\beta\right)^2 \left(\left(\frac{x}{\sqrt{2}}+\frac{1}{\frac{x}{\sqrt{2}}+\beta}\right)^2+\gamma-3\right)\,.
\label{PotPolQ}
\end{equation}
There is a stable Minkowski vacuum for $\beta=2-\sqrt{3-\gamma}$. We saw that without the inclusion of a quintessence scalar the values $0\leq \beta < 2-\sqrt3$ were disallowed by the de Sitter conjecture. For the model in \eqref{PotPolQ} this corresponds to the range $0\leq \beta <2-\sqrt{3-\gamma}$. However notice that now the de Sitter conjecture is satisfied for this range of $\beta$ as long as $\sqrt{2\gamma} \geq c$.
\begin{figure}[ht]
    \centering
    \includegraphics[width=0.75\textwidth]{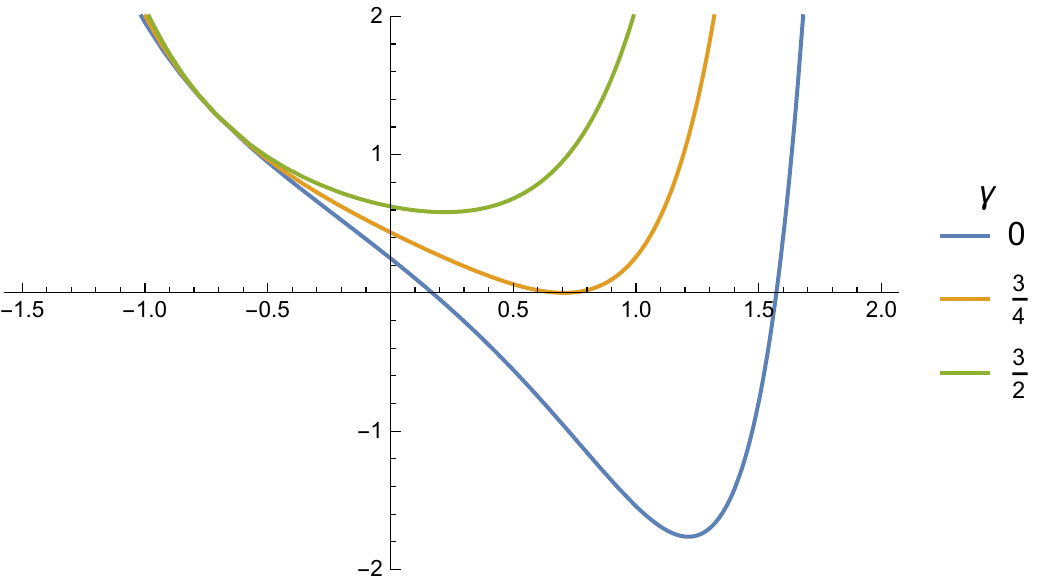}
    \caption{The potential $\widehat{V}(x,0)$ of the Pol\'onyi model with a quintessence scalar that is positioned at $\phi=0$ for $\beta=1/2$ with $\gamma=\left\{0,\,3/4,\,3/2 \right\}$. The addition of the quintessence scalar causes an uplift of the vacuum energy corresponding to the steepness of the potential in the direction of $\phi$.}
    \label{PlotPolQ}
\end{figure}
\bigskip

Until now we only considered the case of single exponential potentials, but the analysis above can be extended to multiple exponential functions and fields by enlarging the K\"ahler and superpotential to
\begin{equation}
\widehat{K}=K-\kappa ^{-2}\sum_{i=1}^{N}a_i \log \left(\Phi_i+\overline\Phi_i\right)\,,\quad \widehat{W}=W\prod_{i=1}^{N}\left(\sum_{j=1}^{M_i}\Phi_i^{-b_{ij}/2}\right)\,.
\end{equation}
The requirement for stable minima in the $z^\alpha$, $\bar z^{\bar\alpha}$ directions will again be an equation of the form in \eqref{cond2}, but with $\gamma(\phi_i)$ generically being a function of the quintessence fields. Therefore in order for these minima to remain fixed during the running of the quintessence scalars, both terms in \eqref{cond2} should vanish independently at the same position. This is a highly restrictive condition, but can be trivially fulfilled when the uplift is caused by a nilpotent superfield(s). 
\vspace{3mm}

\textbf{Example 3.} As a last example one can consider the case where an extra power law is added to \eqref{KWex}:
\begin{equation}
\widehat{K}=X\bar X-a\kappa ^{-2}\log(\Phi+\bar\Phi)\,,\quad W=\left(\Phi^{-b_1/2}+\Phi^{-b_2/2}\right)(m\kappa ^{-2}+fX)\,.
\label{KWex2}
\end{equation}
The scalar potential in this case is given by
\begin{equation}
\widehat{V}=\Lambda_1\,\rme^{\sqrt{2\gamma_1}\phi}+\Lambda_2\,\rme^{\sqrt{2\gamma_2}\phi}+\Lambda_{3}\,\rme^{\left(\sqrt{\gamma_1/2}+\sqrt{\gamma_2/2}\right)\phi}\,,
\label{NilpV2}
\end{equation}
where $\gamma_1=(a+b_1)^2/a$, $\gamma_2=(a+b_2)^2/a$ and
\begin{align}
\Lambda_1&=f^2-\kappa^{-2}m^2\left(3-\gamma_1\right)\,,\nonumber\\
\Lambda_2&=f^2-\kappa^{-2}m^2\left(3-\gamma_2\right)\,,\nonumber\\
\Lambda_3&=f^2-\kappa^{-2}m^2\left(3-\sqrt{\gamma_1\gamma_2}\right)\,.
\label{coefnil}
\end{align}
\newpage

\section{Conclusion}
The de Sitter conjectures (\ref{rdscone}),~(\ref{rdsctwo}), assumed to be imposed by quantum gravity, constrain the class of EFTs belonging to the landscape. Whenever supersymmetry is broken, supergravity theories will not automatically satisfy the conjectures. We derived the constraints on the K\"ahler-invariant ${\cal G}$-function in these cases and explicitly discussed their impact on several models. We studied three types: the Pol\'onyi model, $R$-symmetric one scalar models and models with exponential potentials.

The Pol\'onyi model violates both conjectures in the range $0\leq \beta<2-\sqrt3$. The troublesome range for the R-symmetric one scalar model is $3/4<\alpha\leq 1$ where both the conjectures are also violated. For all other values at least one of the conjectures is satisfied. These parameter ranges might be expanded when the value of constant $c$ is taken into account.

Models with exponential potentials are important for quintessence phenomenology. In this paper we provided a method to obtain exponential potentials in supergravity such that the phenomenological constraint $ \xi \lesssim 0.6$ can be satisfied while also having a positive vacuum energy $V_0>0$. The method can furthermore be used to rescue models from disobeying the de Sitter conjectures by including a quintessence scalar into the theory.

\section*{Acknowledgments}
We would like to thank N. Cribiori, C. Roupec, A. Sagnotti, M. Scalisi and T. Wrase for discussions.  The work of S.F. is supported in part by CERN TH Dept and INFN-CSN4-GSS. The work of M.T. and A.V.P. is supported in part by the KU Leuven C1 grant ZKD1118C16/16/005.  The work of M.T. is supported by the FWO Odysseus grant G.0.E52.14N.


\providecommand{\href}[2]{#2}\begingroup\raggedright\endgroup

\end{document}